\documentclass[a4paper,11pt]{article}
\usepackage{moreverb}
\usepackage[para]{threeparttable}
\usepackage{amssymb,amstext,amsmath}
\usepackage{rotating}
\usepackage{natbib}
\usepackage[T1]{fontenc}
\usepackage{longtable}
\usepackage{graphicx}
\usepackage{times}
\usepackage{setspace}
\usepackage{eucal}
\usepackage{longtable}
\usepackage{enumerate}
\usepackage{float}
\usepackage[caption = false]{subfig}
\usepackage{graphicx}

\def\s{\sigma^2}

\def\m{\mu}
\def\m1{\mu_1}
\def\m0{\mu_0}

\def\cij{_{1,ij}}

\def\qij{_{2,ij}}
\def\ijk{_{ijk}}
\def\ellijk{_{\ell,ijk}}
\def\cijk{_{1,ijk}}
\def\qijk{_{2,ijk}}

\def\ICC{\mbox{ICC}}

\begin{document}


\title{A comparison of multiple imputation methods for bivariate hierarchical outcomes}

\author{K.~DiazOrdaz, M.~G.~Kenward, M.~Gomes, R.~Grieve}

\maketitle



\begin{abstract}
Missing observations are common in cluster randomised trials. Approaches taken to handling such missing data include: complete case analysis, single-level multiple imputation that ignores the clustering,  multiple imputation with a fixed effect for each cluster and multilevel multiple imputation.

We conducted a simulation study to assess the performance of these approaches, in terms of confidence interval  coverage and empirical bias in the estimated treatment effects. Missing-at-random clustered data scenarios were simulated following a full-factorial design. An Analysis of Variance was carried out to study the influence of the simulation factors on each performance measure.

When the randomised treatment arm was associated with missingness, complete case analysis resulted in biased treatment effect estimates. Across all the missing data mechanisms considered, the multiple imputation methods provided estimators with negligible bias. Confidence interval coverage was generally in excess of nominal levels (up to 99.8\%) following fixed-effects multiple imputation, and too low following single-level multiple imputation. Multilevel multiple imputation led to coverage levels of approximately 95\% throughout.

The approach to handling missing data was the most influential factor on the bias and  coverage. Within each method, the most important factors were the number and size of clusters, and the intraclass correlation coefficient.
\end{abstract}


\section{Introduction}
\label{introduction}

In cluster randomised trials, the unit of random allocation is a group of individuals (e.g. a school or a hospital) rather than the individual subjects. It is a common study design in the health and social sciences, especially for evaluations of interventions that operate at a group level,
manipulate the socio-physical environment, or  cannot be delivered
at an individual level.
It is well-known that observations  within each cluster are correlated \citep{Cornfield1978}   and that analyses that ignore this homogeneity within clusters can result in overestimation of the precision of the treatment effects, possibly leading to inappropriate inferences being drawn.  Appropriate statistical techniques for cluster randomised trials are well developed and include mixed models and generalised estimating equations \citep{donner2000}.

A common problem that compromises the validity of the results is that of missing data. The validity of inferences from  incomplete data depends on the process that leads to data being missing, the so-called missing data mechanism, also known as {\it missingness mechanism} or {\it missing data process} \citep[Section 3.2]{mk07}. The missing data mechanism is characterised by the conditional distribution of the probability of missingness, given the data. \cite{rubin1987} proposed a classification of the missing data mechanisms, according to the assumed model for the probability of non-response. A process is said to be Missing Completely at Random (MCAR) if the probability of non-response is completely independent of the measurement process. A process is classified as Missing at Random (MAR) if the probability of non-response is conditionally independent of the unobserved data given the observed data. Processes that are neither MCAR nor MAR are called missing not at random (MNAR).

For missing data mechanisms that satisfy MAR,  valid inferences can be obtained using likelihood-based or Bayesian analyses of the complete cases \cite[Part III]{mk07}. However, moment-based estimators, such as those that use generalised estimating equations are, without special modification, only valid with more stringent conditions about the missing data mechanism, namely that the data are MCAR.

A commonly used approach to obtain valid inferences for incomplete data under the MAR assumption is Multiple Imputation (MI) \citep{rubin1987}. In some circumstances, essentially when the analysis and imputation models coincide,  MI principally replicates a likelihood analysis. However, an advantage of MI is that unlike conventional likelihood analyses, it can incorporate so-called auxiliary variables that are not included in the analysis model,  but which are related to both the missing values and to the probability of observations being missing. Incorporating such auxiliary variables makes the underlying MAR assumption more plausible.

 From a theoretical perspective, it is known that for cluster randomised trials, the imputation method should accommodate the multilevel structure of the data. A failure to do this may lead to invalid inferences \citep{schafer2002}.   However, multilevel multiple imputation is not yet available as a standard implementation in commonly used statistical packages, although particular routines are available, for example, \cite{schafer2001,cgk11}. Hence,
analyses using MI in the cluster randomised trials settings commonly avoid such imputation strategies, and use instead imputation methods that ignore the clustering \citep{DO2014}. An alternative approach that has been previously recommended in the literature is including the cluster as a fixed effect \citep{White2011, Graham2009}. This has the advantage of being easily implemented in widely available MI software.

 Previous simulation-based comparisons of the alternative methods have been presented by \cite{taljaard2008} and \cite{andridge2011}, using a single missing outcome and missing data mechanisms under both MCAR and  MAR dependent on individual-level variables. In the present study, we consider situations where we wish to  simultaneously analyse several responses as functions of the explanatory variables using random effects models. Examples of such models include cost-effectiveness analysis, where policy-makers require an estimate of the alternative treatments on the joint distribution of costs and health outcomes. We focus on bivariate responses with missing data, but the conclusions for univariate or multivariate outcomes follow directly from these. The simulation does not  consider missing covariate data nor data which are missing not at random.

The aim of this paper is two-fold. The first is to investigate the relative performance of different multiple imputation strategies for handling missing bivariate outcome data in cluster randomised trials, over a wide range of missingness mechanisms that are dependent on individual and cluster-level variables. The second is to explore the effect of different trial characteristics, such as number and size of the clusters and level of clustering, on the performance of the MI estimator in finite samples given one of the above MI strategies.  A simulation study with a factorial design is used for this.

The remainder of this paper is organised as follows. In the next section, we
provide some details on the alternative MI methods.  In Section \ref{sec:simulation},
we describe the simulation study and its analysis, in particular making use of the factorial
structure through analysis of variance type procedures. In Section \ref{sec:simresults},
we report  a selection of results from simulated scenarios.  We close with a few points
of interpretation and discussion in Section \ref{sec:discussion}.

\section{Multiple Imputation}\label{sec:missing}

Multiple imputation breaks down the analysis of incomplete data into a number of steps.
We  first need to distinguish between two statistical models. The first
is the analysis model that would have been used had the data been complete. This is called the \textit{substantive model} or \textit{model of interest}. The second model, called the
{\it imputation model},  is used to describe the conditional distribution of the missing data given the observed. For hierarchical data, this conditional distribution must reflect the multilevel nature of the data.

The MI algorithm proceeds by fitting the imputation model to the observed data and taking Bayesian draws from the posterior distribution of its model parameters. Missing data are then imputed from the imputation model, using the parameters previously drawn. These steps are repeated a fixed $M$ number of times, to obtain $M$ {\em completed} data sets. The substantive model is  then fitted to the multiple data sets separately, producing $M$ sets of parameter and covariance estimates which are combined using Rubin's formulae \citep{rubin1987} to produce a single
MI estimate of the substantive model parameters and associated covariance matrix.

Under the MAR assumption, this will produce consistent estimators and, in the absence
of auxiliary variables, is asymptotically (as $M$ increases) equivalent to maximum likelihood \citep{lr02,schafer97}.

Sampling from the approximate predictive distribution of the missing data  as described
above can be performed in several ways. Two broad approaches can be identified; the first approach jointly models incomplete variables, by sampling from an underlying joint predictive distribution \citep{schafer97, gckl09}. In the second approach, referred to as full-conditional specification (FCS) or {\em chained equations},
 draws from the joint distribution are approximated using a sampler consisting of a set of univariate models for each incomplete variable conditional on all the other variables \citep{vb12}.

In the simulations presented here,  both approaches are used. For single-level imputation and fixed cluster effects models, which are also essentially single-level, the FCS method is used, as implemented in the MICE package in R \citep{vb11}. The FCS approach is not well-suited to proper multilevel MI and so, for these imputations, Schafer's PAN package is used \citep{schafer2001}.

Having outlined the generic  MI procedure, we now set out the details of the relevant imputation models to be compared here.

Let $Y_{1,ij}$ and $Y_{2,ij}$ be the two continuous outcomes with missing data, corresponding to the
$i$-th individual in cluster $j$ of a two-arm cluster trial. Let treatment allocation
be represented by $k=1$, if the cluster is allocated to intervention, and 0 otherwise.
Let ${\bf X\ijk}$ denote the matrix of all auxiliary variables (assumed to be fully observed), including individual and cluster-level variables.

The imputation models compared here express
$(Y\cijk, Y\qijk)$  as a function of the grand mean in treatment arm $k$  ($\nu_{\ell,0k}$,
for $\ell\!=\!\{1,2\}$), the auxiliary variables, and error terms $(e\cijk,e\qijk)$.

The single-level imputation model (SMI) can be written as:
\[
\begin{array}{c}
Y\cijk = \nu_{1,0k}+{\bf X\ijk}\nu_{1,X} +e\cijk \\
Y\qijk = \nu_{2,0k}+{\bf X\ijk}\nu_{2,X} +e\cijk
\end{array}\ \ \ \ \ \ \ \ \begin{array}{c}
\left(\!\!\begin{array}{c}
e\cijk\\ e\qijk\end{array}\!\!\right)
\sim \mbox{N}(\bf{0},\bf{\Omega_1})\end{array}
\]
where $\nu_{\ell, X}$ is the vector of regression coefficients,
and $\bf\Omega_1$ is the individual-level variance-covariance matrix.
With single-level MI, the imputed values are drawn from the conditional distribution
of the missing observations given the observed data, ignoring any dependency between
observations within a cluster not explained by the cluster-level auxiliary variables
included in the model. Therefore, the single-level imputation model does not properly
represent the conditional distribution of the missing data given the observed data.

Two imputation models have been used to incorporate the effect of clustering.
Firstly, we include a cluster fixed-effect in the imputation model (denoted FMI):
\begin{eqnarray*}\label{fixed}
Y\cijk &=& \nu_{1,0k}+{\bf X\ijk}\nu_{1,X} +\sum_{j=1}^{J-1}\beta_{1,j} I_{ij}+e\cijk \\
Y\qijk &=& \nu_{2,0k}+{\bf X\ijk}\nu_{2,X} +\sum_{j=1}^{J-1}\beta_{2,j} I_{ij}+e\qijk
\end{eqnarray*}
where $I_{ij}$ is the indicator variable for cluster $j$, so that $I_{ij}=1$ if the observation $i$ belongs to cluster $j$ and the error term $(e\cijk,e\qijk)$ is assumed to be bivariate normal as before.
This model allows a different intercept for each cluster within treatment group $k$.
Missing outcomes will be imputed from the conditional normal distribution
given the other outcome, if observed, and the auxiliary variables, which
must all be at the individual level, with a mean determined by the fixed-effect for that cluster.

Secondly, we include a {\it random effects} for clustering in the imputation model (denoted MMI):
\[
\begin{array}{c}
Y\cijk = \nu_{1,0k}+{\bf X\ijk}\nu_{1,X}+b_{1,j}+e\cijk \\
Y\qijk = \nu_{2,0k}+{\bf X\ijk}\nu_{2,X}+b_{2,j}+e\qijk
\end{array}\ \ \ \ \ \ \ \ \begin{array}{c}
\left(\!\!\begin{array}{c}
b_{1,j}\\ b_{2,j}\end{array}\!\!\right)
\sim \mbox{N}(\bf{0},\bf{\Omega_2})\end{array}
\]
where $\bf\Omega_2$ is the cluster-level variance-covariance matrix and
the  individual-level residuals $(\!e\cijk,e\qijk\!)$ are assumed normally
distributed independently of $(b_{1,j}, b_{2,j})$, the cluster random-effects.

Finally, complete-case analysis (CCA) is also included in our simulations, for comparative purposes.

\section{Simulation study}\label{sec:simulation}

A simulation study following a full factorial design was conducted comparing the performance of the methods considered here. The simulation steps proceeded as follows: data generation, application of a missing data mechanism, and
estimation and inference for the treatment effect from the analysis after handling (or ignoring) the missing data. Finally, the behaviour of the treatment
effect estimator is examined according to our chosen performance measures.

\subsection{Data generation}

For each subject $i$ in cluster $j$,   standard normal individual-level covariate $X_i$
and cluster-level variable $W_j$  were generated. Bivariate normal outcome data $(Y\cijk,Y\qijk)$ were then generated depending on these covariates, separately in each treatment arm $k=0,1$.

The level of clustering, quantified by the intraclass correlation coefficient (ICC),  was allowed to vary according to the levels set out in Table \ref{simfactors}. The number and size of clusters were also varied, while maintaining the same overall sample size ($S=500$).
Three different types of cluster randomised trial design were considered: (i) large number of clusters ($J=50$) and few individuals per cluster ($n_j=10$); (ii) small number of clusters ($J=10$) and large cluster size ($n_j=50$);
(iii) moderate number of clusters (30) and variable number of individuals per cluster. For these scenarios, cluster size $n$  was assumed to follow a Gamma distribution, with mean 20 and coefficient of variation $cv=\frac{\textit{SD(n)}}{\textit{E(n)}}=0.5$.

\subsection{Missing data mechanisms}

To generate the missing data under the Missing-at-Random assumption, we used four different missing data mechanisms, where the probability of non-response, denoted by $\pi_{\ell,ijk}$, was such
that the non-response indicator $R_{\ell,ijk}\sim \textit{Bern}(\pi_{\ell,ijk})$, depends on $X_i$ or/and $W_j$, as displayed also in Table \ref{simfactors}. The coefficient $\eta$ represents the strength of association between the covariates and non-response indicator $R_{\ell,ijk}$.We adjusted $\alpha_0$ empirically to achieve the required expected probability of missing.

We selected other factors which were anticipated to have an impact on the performance of the approaches for handling missing data, based on previous literature \citep{rubin1987, taljaard2008, andridge2011, ck13}.
For the first three missing data mechanisms, the same factors and levels were used for both randomised treatment arms. These are reported in Table \ref{simfactors}.

 We assumed that for both outcomes, individual and cluster level covariates have the same level of association $\eta$ with the non-response indicator, and thus we drop the indexes $\ell$ and $X, W$.  However, for the last of our missingness mechanisms, we allowed $\eta$ to differ between treatment arms, with two settings, and these are presented in Table \ref{simfactorsdifftrt}, together with the probabilities of non-response, which also differ across treatment arms.

Non-response rates were chosen to minimise the number of clusters with
one or both outcomes completely missing, as whole cluster non-response raises other issues not dealt here.

For each simulated dataset, non-response indicators $R\ellijk$ for each outcome were independently drawn from a Bernoulli distribution with probabilities $\pi\ellijk$ as specified in Table \ref{simfactors}. Missing values were then generated to create the {\it observed} data set.

\subsection{Substantive model}\label{sec:substantive}

We focus on likelihood-based methods for the substantive model,  rather than estimating equations. Because the intervention effect lies at the cluster level, likelihood-based methods that acknowledge the clustering must use \textit{random} cluster effects; fixed cluster effects would absorb all the information on the intervention effects.

Hence, the substantive model is a bivariate Gaussian random-effects model where the only explanatory variable is treatment.
Let the cluster-level random effects be represented by the latent variables $u_{1,j}$ and $u_{2,j}$.  The model can be written as follows
\begin{eqnarray}\label{anmodel}
\nonumber Y\cij&=&\beta_{1,0}+\beta_{1}t_j+u_{1,j}+e\cij\\
Y\qij&=&\beta_{2,0}+\beta_{2}t_j+u_{2,j}+e\cij
\end{eqnarray}
where $\beta_{1}$ and $\beta_2$ represent the treatment effect on the corresponding outcome.
The error term $(e\cij,e\qij)$ and the cluster effects are assumed to be normally distributed:
\[
\left(\!\!\begin{array}{c}
e\cij\\ e\qij\end{array}\!\!\right)
\sim
\mbox{N}\left[\!\left(\!\begin{array}{c} 0\\ 0\end{array}\!\right),
\left(\!\!\begin{array}{c c}\s_{1} & \rho\sigma_1\sigma_2 \\ \rho\sigma_1\sigma_2 & \s_{2}\end{array}\!\!\right)
\!\right] \mbox{\ and\ }
\left(\!\!\begin{array}{c}
u_{1,j}\\u_{2,j}\end{array}\!\!\right)
\sim
\mbox{N}\left[\!\left(\!\begin{array}{c} 0\\ 0\end{array}\!\right),
\left(\!\!\begin{array}{c c}\tau^2_{1} & \phi\tau_1\tau_2 \\ \phi\tau_1\tau_2 & \tau^2_{2}\end{array}\!\!\right)
\!\right]
\]
where  $\sigma_1, \sigma_2$ are the individual-level standard errors, $\rho$ is the
individual-level correlation between $Y_1$ and $Y_2$ and  $\tau_1, \tau_2,$ and $\phi$
 are the standard errors and correlation of the two cluster random effects, respectively.

\subsection{Implementation}

The number of imputations $M$ was set at 10. After imputation, for which the two covariates were used as auxiliary variables, the substantive model,
equation (\ref{anmodel}), was applied to each multiply imputed dataset to estimate treatment effect on $Y_1$ and $Y_2$ simultaneously. The estimates obtained using the analysis model in each of the $M$  multiply imputed sets were then
combined using Rubin's rules.

For each scenario, the whole simulation procedure (data generation, imposing missing values, imputation, analysing each of the imputed datasets using the substantive model, and combining the resulting treatment effect estimates using Rubin's rules) was performed on $N=1000$ datasets to capture the behavior in repeated samples.

\subsection{Performance criteria}

Let $\theta$ denote the true treatment effect parameter, and  $\hat{\theta}_l$ the estimate obtained in the $l=1, \ldots,N$ replicated dataset.
The following criteria were used to measure the  performance of the different MI strategies.

\begin{enumerate}
\item Confidence interval coverage rate (CR): The percentage of times that the true parameter value is covered in the 95\% confidence interval.
\item Empirical bias, $B=\frac{1}{N}\sum_{l=1}^{N}\hat{\theta}_l-\theta$
\item Root-mean-square error (RMSE) $\sqrt{\frac{1}{N}\sum_{l=1}^{N}(\hat{\theta}_l-\theta)^2}$
\item Average width of confidence interval (AW): The distance between the average lower and upper confidence interval limits across $N$ confidence intervals.
\end{enumerate}

The performance of a procedure is regarded as poor if its coverage drops below 90\% \citep{Collins2001}.  If the procedure results in CRs that are close to 100\% extra caution should be taken when using that procedure \citep{yucel2010}.  A CR close to the nominal value, along with narrow confidence intervals translates into greater accuracy and higher power.

\subsection{Analysis of the simulation results}

The factorial structure of the simulation design was exploited through the use of analysis of variance (ANOVA) summaries to isolate key factors that are associated with large impact on the performance of the multiple imputation  procedures.

 ANOVA was carried out on each performance measure for each outcome including all main effects and interactions amongst factors up to 4-way interactions (the rest constituting the ``residual'' degrees of freedom). The relative size of the F-statistics derived from the ANOVA was used as an indicator of the influence of that factor (or interaction of factors) on the particular performance measure. The F-statistic can be
 thought of as the ratio of variability explained/variability unexplained by the model.
Multivariate ANOVA (MANOVA) was also used to study the impact of each factor on the overall performance. For this, the Wilks Lambda statistic was calculated to obtain an approximate F-statistic.

These F-statistics are not used in an inferential way, and no distributional assumptions are being made; instead they are used as a descriptive measure of influence.
To help visualise this, we normalised the value of these F-statistics by dividing each F-value by the largest of those obtained in for each performance measure. For bias and confidence interval coverage, we calculated the proportion of simulated scenarios
where the performance measure in turn was deemed unsatisfactory, that is bias which is larger than 1.96 times the Monte Carlo error, and confidence interval coverage which is either lower than 90\% or higher than 97\%. We then multiply the re-normalised F-values by this proportion and plotted  the resulting number by performance measure and missing data approach.

Since one of our aims is to establish what influences performance within each MI method, these analyses were also performed stratified by MI method.
In addition, we report the range of the percentage bias and coverage rate over each of the simulation factors
and plot the distribution of these performance measures stratified by MI method and missing data mechanism.

\section{Results}\label{sec:simresults}

Without loss of generality, we present here the results corresponding to bias and coverage for treatment effect estimates on $Y_1$. The corresponding results for $Y_2$ are available from the corresponding author upon request.

\begin{figure}\thispagestyle{empty}
\caption{Boxplot of the distribution of (a) percentage bias and (b) coverage rate for treatment effect estimates on $Y_1$, by analysis strategy
(CCA, SMI, FMI, MMI) and missingness mechanism (the columns denoted by  Ind: individual covariate; Clus: cluster-level
covariate, Both and Treat: indicating the variables associated with missingness). The dotted black lines represent (a) no bias and (b) the nominal coverage rate, while the dashed lines represent minimum and maximum acceptable coverage rates.}
\label{fig_boxplots}
\subfloat[Distribution of Bias]{\includegraphics[width = 5.5in]{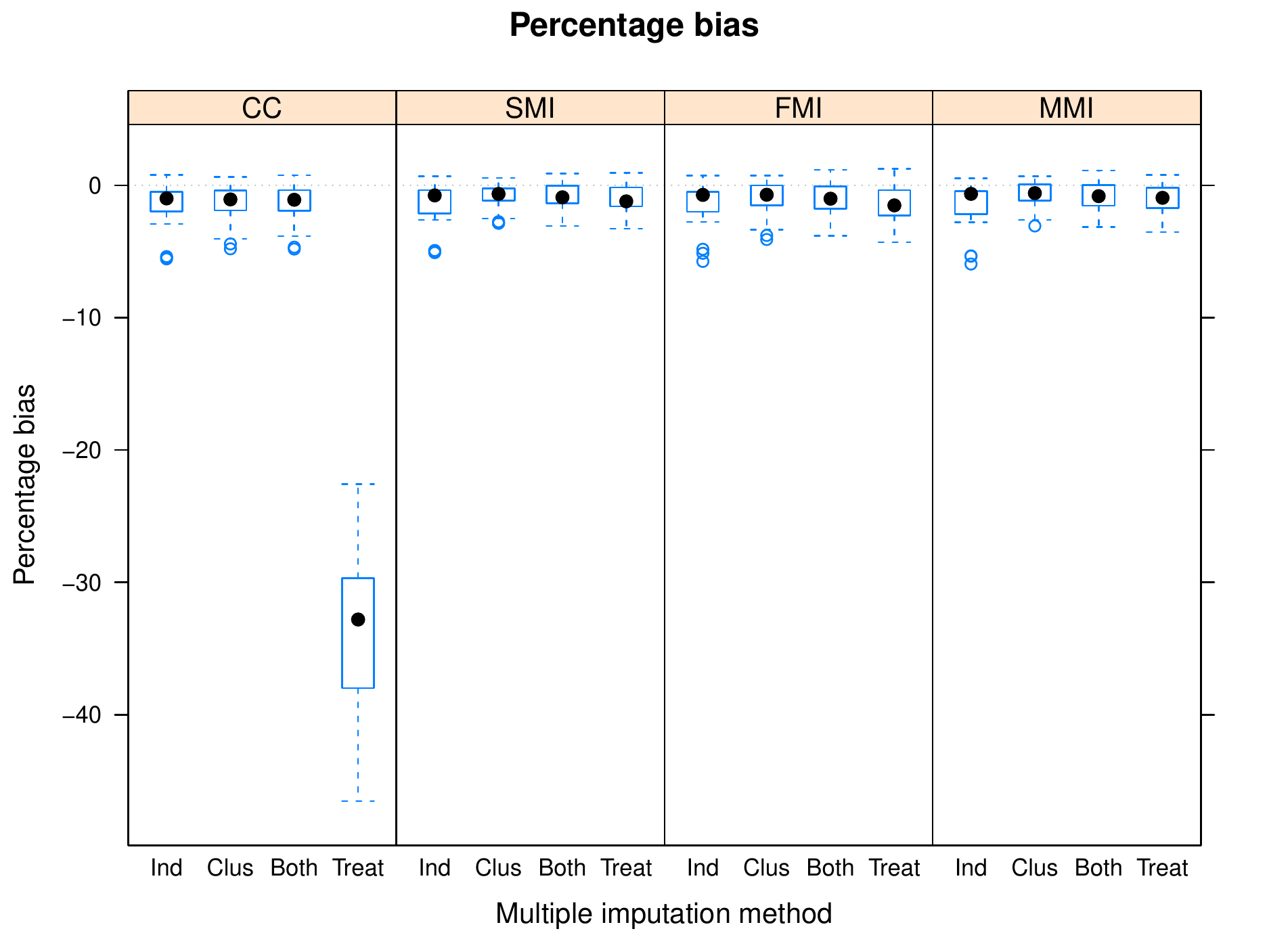}}\\
\subfloat[Distribution of Coverage rate]{\includegraphics[width = 5.5in]{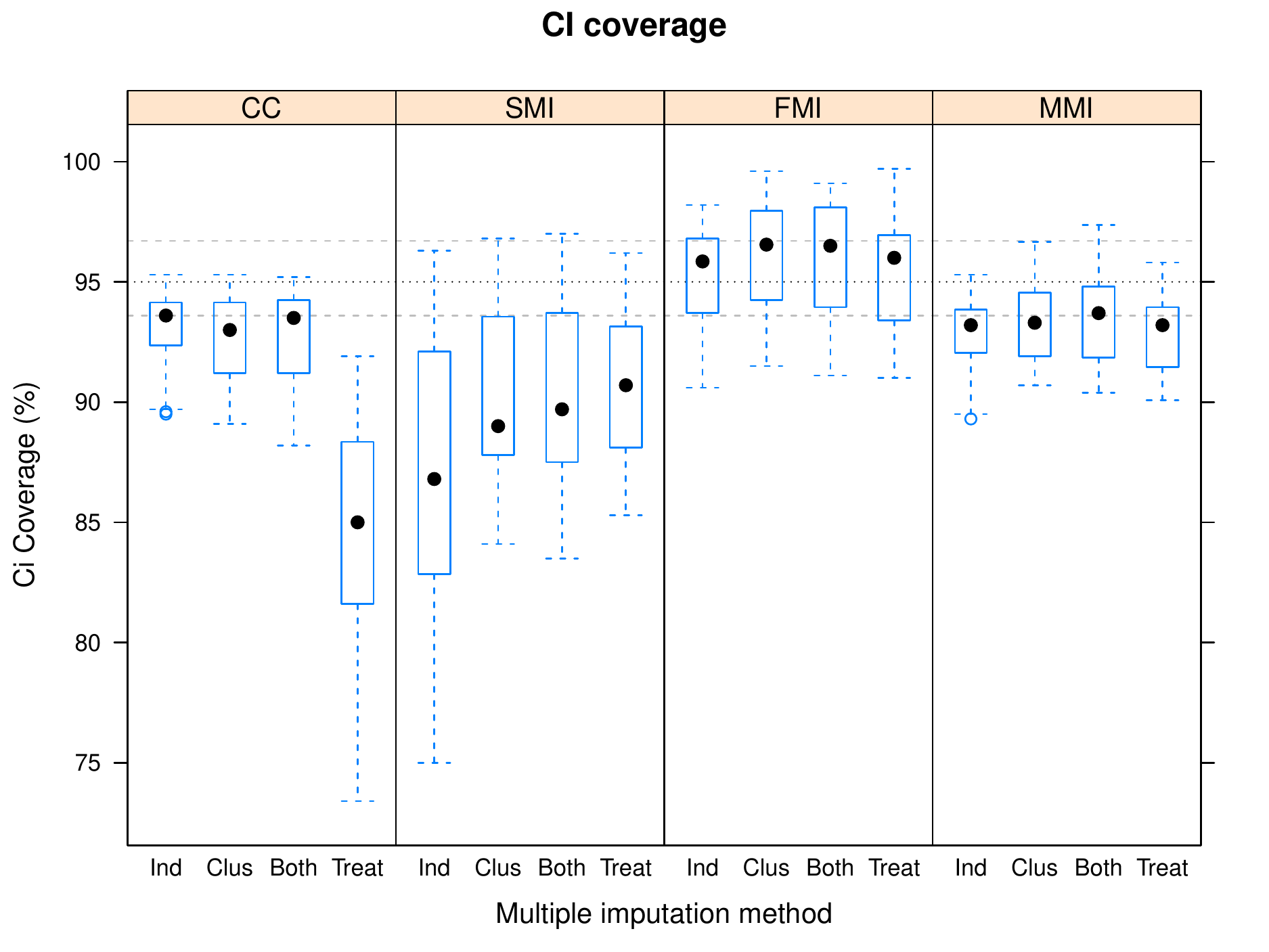}}
\end{figure}

The distribution of bias and coverage rate by MI method and missing mechanism are shown in Figure \ref{fig_boxplots}.  We observe that, for the first three missing data mechanisms studied (see Table \ref{simfactors}), where the missing data mechanism was not dependent on treatment arm, all approaches resulted in unbiased estimates across most of the scenarios. This is in line with theoretical results,  as the variables associated with missingness are not associated with the treatment effect. However, for the scenario when the missing mechanism is differential by treatment arm, the CCA produced substantially biased estimates across the scenarios considered. The corresponding results
for the MI estimates show none to negligible bias: in general less than 3.5\%, and mostly within Monte Carlo error limits. This is reported in Table \ref{tab:biastrt}.

However, the alternative MI strategies resulted in very different variance estimates, and coverage rates. Table \ref{tab:covaw} reports the coverage rate and average width again for the scenarios where the missingness mechanism was different by treatment arms. Similar tables corresponding to the other missingness mechanism are reported in the Supplementary file.
In particular, the single-level MI resulted in substantial under-coverage for scenarios with high ICCs (0.20 and above).
The number and size of clusters also appear to be factors associated with low coverage rate.
For scenarios where the number of clusters is relatively low ($J=5$ per arm), multilevel and single-level MI both have low coverage for the estimated treatment effect.
Fixed-effects MI results in over-conservative coverage for a range of scenarios, especially those corresponding to missing data mechanisms that depend only on a cluster-level variable (Table 5 in the Supplementary File) and where the ICCs are moderate to small.
 Across all methods and scenarios considered, the accuracy is similar, i.e. comparable RMSE, however FMI is
 systematically inefficient, wider confidence intervals are obtained using FMI compared to those obtained using either SMI or MMI, even when estimates are unbiased and coverage is acceptable.

\begin{figure}
\caption{ANOVA results: The height of the bar represents the (normalised) F-statistic value, scaled according to the proportion of scenarios where (a) bias was larger than $1.96 \times$ the Monte-Carlo error, and (b) those scenarios where either under or over-coverage are an issue}\label{anova_biascover}
\subfloat[Bias]{\includegraphics[width = 5.5in]{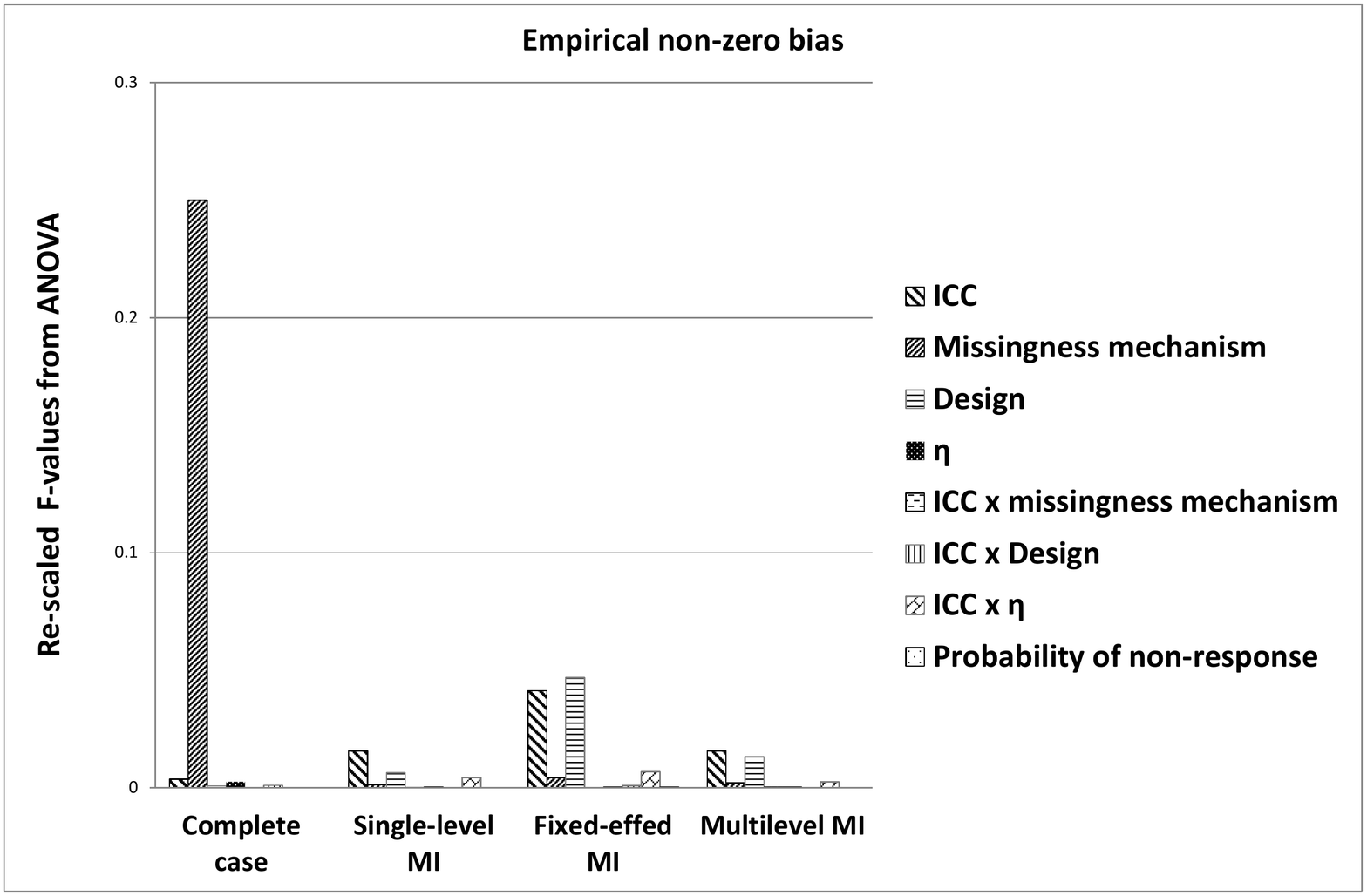}} \\
\subfloat[Coverage rate]{\includegraphics[width = 5.5in]{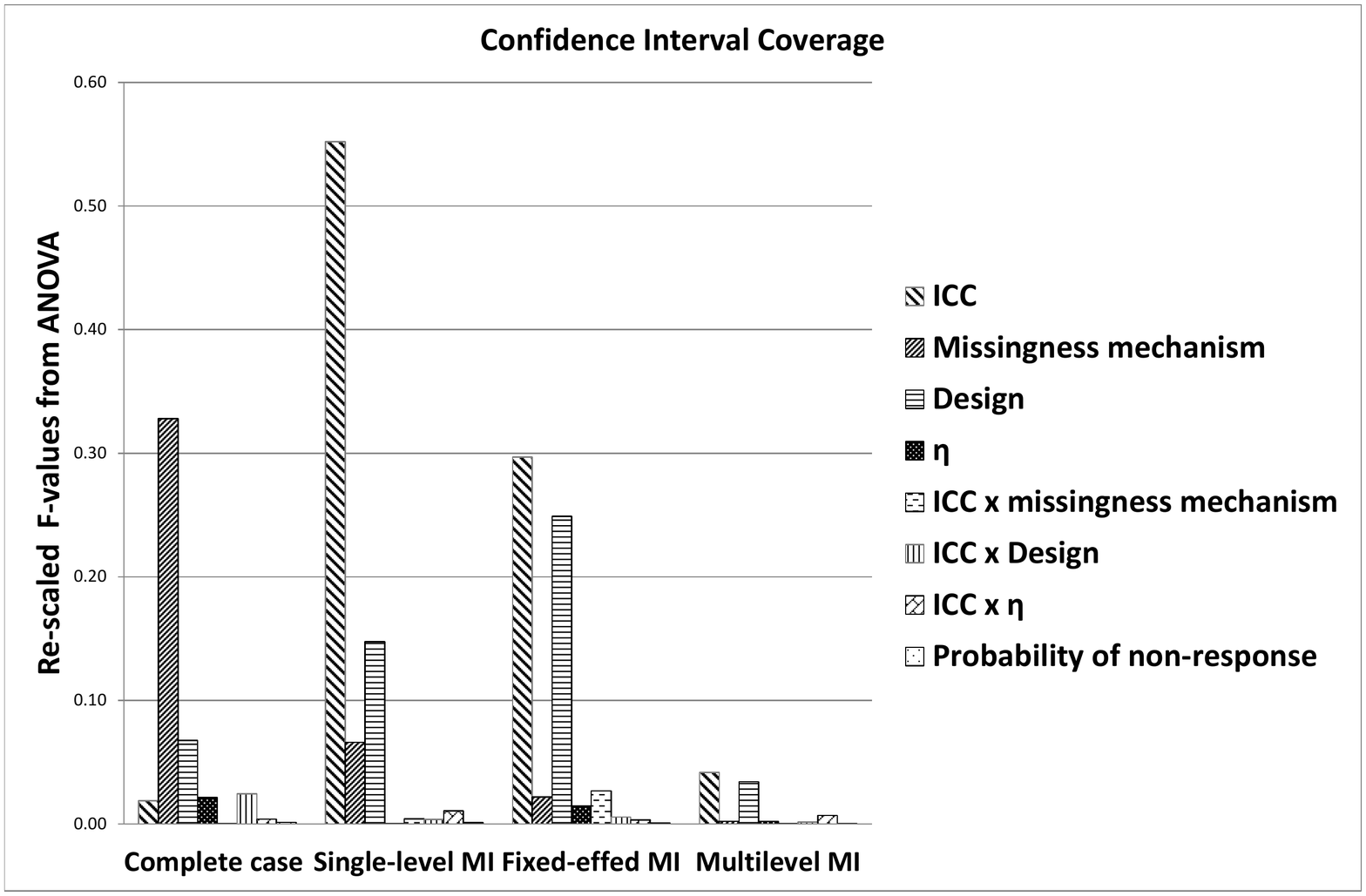}}
\end{figure}

The ANOVA results confirm that the most influential factor on the validity of inferences drawn from missing data is the method chosen to handle these.
Within each MI method, the ANOVA results on bias and coverage rate are reported graphically in Figure \ref{anova_biascover}. Terms with very small normalised F-statistic value are not plotted.
For CCA, the most influential factor for the substantial empirical bias is the missing data mechanism.  For SMI and MMI empirical bias is low, and the
strength of association between the covariates and the
non-response indicator is almost as influential as the missing
data mechanism. For FMI, while bias is again low, the ANOVA suggests that ICC
and the number and size of the clusters are the
determining factors affecting bias.

The corresponding figure for coverage rate shows that most influential factors are the level of clustering,  measured by the ICC and the
 number and size of the clusters (denoted in the figures as {\em Design}). Nevertheless,
 for CCA, the most influential factor is the
 missing data mechanism.
As the relative height of the bars show, the method which
 most consistently achieves coverage rates
 close to the nominal is MMI, as it has the
 smallest proportion of scenarios with over
 or under-coverage, with only 8 scenarios out of the total 192 resulting in CR lower than 90\% and higher than 97\%.

\section{Discussion}\label{sec:discussion}

In this simulation study, we compared the performance of single, multilevel and fixed-effects MI for handling missing data in cluster randomised trials. The full-factorial nature of our simulation study enabled us to establish which characteristics have the greatest influence on the performance of the alternative methods for handling missing data considered here.

In our simulations, which assumed the data were MAR throughout, bias was a serious problem for the complete case analysis when the missingness mechanism was differential by treatment arm, while all MI methods resulted in unbiased treatment estimates.
The main difference amongst the three MI procedures is in how variability is incorporated into the imputations. Single-level MI resulted in low ($<90\%$) coverage rate across most scenarios, in particular when the ICCs exceeded 0.05 and there were few clusters. Fixed-effects MI produced overly conservative coverage ($>97\%$), especially when there were small ICCs and more than 30 clusters. This finding reflects the way these two approaches accommodate the between-cluster variance. Under single-level MI, the between-cluster variance is set to zero, whereas with the fixed-effect MI this variance is unbounded in the sense that the behaviour of one estimated cluster effect is unrelated, or unconstrained, by the behaviour of any of the others.  Indeed, including cluster as a fixed-effect in the imputation model represents the limiting case where the proportion of variability at the cluster-level tends to one and does not properly capture the conditional distribution of the missing data given the observed. It cannot be used  when cluster-level variables need to be imputed, and appears to perform worse when the missing data mechanism is driven by a cluster-level covariate, which cannot be explicitly included in the imputation model.

By contrast,  multilevel MI models the correlation in the data appropriately, producing coverage  rates close to the nominal level.  This consistent performance across the varying sample sizes is indicative of acceptable finite sample properties.  Moreover, multilevel MI is compatible with the substantive model, which uses cluster random effects, and the imputation model can include auxiliary variables at both the individual and the cluster-level, thus increasing the plausibility of the MAR assumptions.

Our findings underscore the importance of selecting an imputation model with a compatible, multilevel structure to that of the substantive models,
and corresponds to those from previous studies which have compared single and multilevel MI in settings with hierarchical data. For example, \cite{taljaard2008} found that single-level MI results in excessive Type I errors in settings where data were MCAR.
Our study also extends the results  of \cite{andridge2011}, who found that including cluster as a fixed effect in the imputation model overestimates the variance, especially when ICCs are low, and there are few clusters.

The results presented in this study could potentially be extended to other situations. Our imputation and substantive models match exactly the data generating process, but previous simulation studies  \citep{schafer97,yucel2010} have shown that MI is fairly robust to distributional misspecification of the imputation model.  Also, for simplicity, we assumed the missing data mechanism is MAR throughout. An interesting extension would be to explore MNAR mechanisms, especially those when the cluster random effect is driving the missingness. Other potential extensions relate to situations where there is cluster non-response. In both situations, multilevel MI could provide a flexible route for investigating sensitivity to alternative MNAR mechanisms and cluster drop-out  \cite[Chapter 10]{ck13}.

\section*{Acknowledgments}
The authors will like to thank James Carpenter for helpful discussions.
KDO is funded by an MRC Career development award in Biostatistics, MG by an MRC Early Career fellowship in Economics of Health, and RG by a senior research fellowship from NIHR.

\clearpage
\section*{Tables and Figures}

\begin{table}[h]
\caption{Factors and their chosen levels that differ across scenarios for missingness mechanisms which do not differ by treatment arm}\label{simfactors}
\begin{tabular}{lll}
    \hline\hline Factor & Levels& Values\\ \hline
$\ICC_1$ and $\ICC_2$ &low&(0.01, 0.01)\\
&moderate&(0.20, 0.05)\\
&high& (0.20, 0.20)\\
&differential by outcome&(0.60, 0.01) \\ \hline

Cluster design&many small clusters &$J=50$, $n_j=10$\\
&few large clusters&$J=10$, $n_j=50$ \\
&unbalanced&$J=30$ , variable size\\ \hline
Missingness
&Individual covariate &$\mbox{logit}\  \pi_{\ell,ij}=\alpha_0+\eta X_i$\\
mechanism&Cluster covariate &$\mbox{logit}\  \pi_{\ell,ij}=\alpha_0+\eta W_j$\\
&Both &$\mbox{logit}\  \pi_{\ell,ij}=\alpha_0+\eta X_{i}+\eta W_{j}$\\
&Differential by treatment &$\mbox{logit} \pi_{\ell,ijk}=\alpha_{0k}+\eta_{k} X_{ij}+\eta_{k} W_j$
\\ \hline
Association between covariates& low& $\eta=1$\\
 and missingness &high&$\eta=2$\\ \hline
Probability of Non-response &equal&$20 \%$ \\
&different by outcome&$30\%$ for $Y\cij$; $10\%$ for $Y\qij$\\\hline
 \hline
\end{tabular}
\end{table}

\begin{table}
\begin{center}
\caption{Parameter values for settings where the missingness mechanism is differential by treatment arm.}\label{simfactorsdifftrt}
\begin{tabular}{llllll}
\hline\hline
&       & Association      & \multicolumn{3}{c}{Probability of non-response} \\
&       &with        &equal   &\multicolumn{2}{c}{Different by outcome } \\
Level of association &  Arm & missingness  &  & For $Y_1$ & For $Y_2$\\  \hline

    low   & Control & $\eta_0=1$ & 20\% &   30\%  & 10\%  \\
          & Intervention & $\eta_1=2$ & \textit{35\%} &  \textit{45\%} & \textit{20\%} \\\hline
    high  & Control  & $\eta_0=1.5$ & 10\% &     15\%  &10\% \\
          & Intervention & $\eta_1=3$ &\textit{30 \%}     &\textit{35\%} &\textit{30\%} \\
 \hline\hline
\end{tabular}
\end{center}
\footnotesize{The numbers in italics are not simulation parameters, but the
approximate empirical rates of non-response obtained after setting $\alpha_0$.}
\end{table}

\clearpage

\begin{table}
\thispagestyle{empty}
 \label{tab:biastrt}
  \centering
  \caption{Percentage bias for the estimated treatment effect on $Y_1$ for scenarios corresponding to missingness mechanism is differential by treatment}
  {\scriptsize
    \begin{tabular}{llllrrrr} \hline
    Design & $\eta$   & Missingness & ICC   & CCA & SMI   & FMI   & MMI \\ \hline
 $J=50$, $n_j=10$ & Low   & .20,.20 & 0.01, 0.01 & -24.8 & -1.4  & -0.8  & -0.8 \\
          &       &       & 0.20, 0.05 & -32.9 & -1.5  & -1.3  & -1.0 \\
          &       &       & 0.20, 0.20 & -33.1 & -1.6  & -1.3  & -1.0 \\
          &       &       & 0.60, 0.01 & -38.7 & -1.4  & -2.1  & -1.4 \\
          &       & .30,.10 & 0.01, 0.01 & -23.2 & -1.3  & -0.7  & -0.2 \\
          &       &       & 0.20, 0.05 & -31.0 & -1.6  & -1.7  & -0.3 \\
          &       &       & 0.20, 0.20 & -31.1 & -1.6  & -1.7  & -0.4 \\
          &       &       & 0.60, 0.01 & -35.9 & -1.8  & -3.5  & -0.5 \\
          & High  & .20,.20 & 0.01, 0.01 & -28.2 & -1.7  & -2.3  & -1.7 \\
          &       &       & 0.20, 0.05 & -37.5 & -1.7  & -3.0  & -2.0 \\
          &       &       & 0.20, 0.20 & -37.9 & -1.9  & -3.0  & -1.9 \\
          &       &       & 0.60, 0.01 & -43.4 & -1.5  & -4.2  & -2.6 \\
          &       & .30,.10 & 0.01, 0.01 & -29.2 & -1.3  & -1.1  & -1.5 \\
          &       &       & 0.20, 0.05 & -39.3 & -1.5  & -2.3  & -1.7 \\
          &       &       & 0.20, 0.20 & -39.7 & -1.6  & -2.3  & -1.7 \\
          &       &       & 0.60, 0.01 & -46.5 & -1.4  & -4.3  & -2.1 \\ \hline
 $J=10$, $n_j=50$ & Low   & .20,.20 & 0.01, 0.01 & -25.2 & 0.1   & -0.2  & -0.9 \\
          &       &       & 0.20, 0.05 & -31.1 & -1.0  & -1.5  & -1.8 \\
          &       &       & 0.20, 0.20 & -31.5 & -1.0  & -1.5  & -1.7 \\
          &       &       & 0.60, 0.01 & -32.7 & -2.8  & -3.7  & -3.5 \\
          &       & .30,.10 & 0.01, 0.01 & -24.5 & -0.1  & -0.4  & -0.9 \\
          &       &       & 0.20, 0.05 & -30.2 & -1.3  & -1.7  & -1.9 \\
          &       &       & 0.20, 0.20 & -30.7 & -1.3  & -1.7  & -1.9 \\
          &       &       & 0.60, 0.01 & -31.7 & -3.3  & -4.0  & -3.5 \\
          & High  & .20,.20 & 0.01, 0.01 & -29.2 & 0.0   & -0.4  & -0.4 \\
          &       &       & 0.20, 0.05 & -36.5 & -1.1  & -1.7  & -1.3 \\
          &       &       & 0.20, 0.20 & -37.0 & -1.2  & -1.7  & -1.2 \\
          &       &       & 0.60, 0.01 & -38.5 & -2.9  & -4.0  & -3.2 \\
          &       & .30,.10 & 0.01, 0.01 & -31.1 & 0.3   & -1.5  & -0.1 \\
          &       &       & 0.20, 0.05 & -38.9 & -0.7  & -1.7  & -0.9 \\
          &       &       & 0.20, 0.20 & -39.5 & -0.8  & -1.6  & -0.7 \\
          &       &       & 0.60, 0.01 & -41.0 & -2.3  & -4.3  & -2.2 \\ \hline
   $J=30$, unbalanced     & Low   & .20,.20 & 0.01, 0.01 & -23.1 & 0.9   & 1.2   & 0.4 \\
          &       &       & 0.20, 0.05 & -30.3 & 0.4   & 0.6   & 0.0 \\
          &       &       & 0.20, 0.20 & -30.6 & 0.3   & 0.5   & -0.1 \\
          &       &       & 0.60, 0.01 & -33.7 & -0.6  & -0.8  & -1.1 \\
          &       & .30,.10 & 0.01, 0.01 & -22.6 & 0.5   & 1.2   & 0.4 \\
          &       &       & 0.20, 0.05 & -29.6 & -0.4  & 0.2   & -0.2 \\
          &       &       & 0.20, 0.20 & -29.8 & -0.5  & 0.3   & -0.3 \\
          &       &       & 0.60, 0.01 & -33.0 & -2.0  & -1.5  & -1.2 \\
          & High  & .20,.20 & 0.01, 0.01 & -26.8 & 0.8   & 0.4   & 0.5 \\
          &       &       & 0.20, 0.05 & -35.4 & 0.3   & -0.3  & 0.0 \\
          &       &       & 0.20, 0.20 & -35.8 & 0.2   & -0.3  & -0.1 \\
          &       &       & 0.60, 0.01 & -39.6 & -0.5  & -1.8  & -1.1 \\
          &       & .30,.10 & 0.01, 0.01 & -28.3 & 0.7   & 0.5   & 0.8 \\
          &       &       & 0.20, 0.05 & -37.6 & -0.2  & -0.6  & 0.3 \\
          &       &       & 0.20, 0.20 & -38.1 & -0.4  & -0.7  & 0.1 \\
          &       &       & 0.60, 0.01 & -42.6 & -1.8  & -2.7  & -0.8 \\ \hline
\end{tabular}%
}
 \end{table}%

\clearpage
\begin{table}\label{tab:covaw}%
\thispagestyle{empty}
         \centering
  \caption{\small{Coverage rate (CR) and average width (AW) corresponding to confidence interval of the treatment effect estimate, when  missingness is differential by treatment arm.}}
  {\scriptsize
  \begin{tabular}{llllrrrrrrrr}
   Design & $\eta$   & Missingness & ICC     &\multicolumn{2}{c}{CCA} & \multicolumn{2}{c}{SMI}   &\multicolumn{2}{c}{FMI}   &\multicolumn{2}{c}{ MMI} \\
&&&&CR & AW& CR & AW&CR & AW&CR & AW\\
\hline
    $J=50$,   & Low   & .20,.20 & 0.01, 0.01 & {\bf81.2}   & 18.6  & 95.5   & 17.9  & {\it98.8}   & 22.3  & 94.9   & 17.5 \\
         $n_j=10$  &       &       & 0.20, 0.05 & {\bf84.7}   & 27.6  & 92.6   & 26.2  & 96.5   & 31.2  & 93.1   & 27.4 \\
          &       &       & 0.20, 0.20 & {\bf 83.9}   & 27.7  & {92.8}   & 26.3  & 96.3   & 31.2  & 93.1   & 27.2 \\
          &       &       & 0.60, 0.01 & {91.3}   & 56.0  & {90.7}   & 51.4  & 95.1   & 58.7  & 94.4   & 56.9 \\
          &       & .30,.10 & 0.01, 0.01 & {\bf82.2}   & 18.7  & 95.5   & 19.4  &{\it 99.7}   & 26.1  & 94.9   & 18.9 \\
          &       &       & 0.20, 0.05 &{\bf 85.2}   & 27.7  &{92.5}   & 26.9  & {\it97.5}   & 34.0  & 93.0   & 28.2 \\
          &       &       & 0.20, 0.20 & {\bf84.6}  & 27.7  & 92.6   & 27.0  & {\it97.5}   & 34.0  & 92.4   & 27.9 \\
          &       &       & 0.60, 0.01 & {91.7}   & 56.0  & {90.7}   & 50.8  & 95.2   & 60.1  & 93.5   & 57.5 \\
          & High  & .20,.20 & 0.01, 0.01 &{\bf 75.1}   & 17.4  & 95.6   & 17.3  & {\it98.7}   & 21.4  & 95.6   & 17.2 \\
          &       &       & 0.20, 0.05 & {\bf81.2}   & 26.7  & {91.8}   & 26.2  & 96.8   & 30.6  & 93.8   & 27.4 \\
          &       &       & 0.20, 0.20 & {\bf81.2}   & 26.6  & {91.9}   & 26.2  & 96.8   & 30.6  & 93.7   & 27.2 \\
          &       &       & 0.60, 0.01 & {\bf89.7}   & 55.2  &{91.0}   & 52.3  & 94.6   & 58.4  & 94.4   & 56.9 \\
          &       & .30,.10 & 0.01, 0.01 & {\bf73.7}   & 17.8  & 96.2   & 18.5  & {\it99.1}   & 23.4  & 95.8   & 18.1 \\
          &       &       & 0.20, 0.05 & {\bf78.4}   & 26.8  & {92.2}   & 26.8  & 96.8   & 32.0  & 93.2   & 28.0 \\
          &       &       & 0.20, 0.20 & {\bf78.6}  & 26.8  &{92.3}  & 26.8  & 96.8   & 32.1  & 93.6   & 27.8 \\
          &       &       & 0.60, 0.01 &{\bf 89.3}   & 55.3  & {90.4}   & 52.2  & 95.0   & 59.1  & 94.4   & 57.4 \\ \hline
     $J=10$,      & Low   & .20,.20 & 0.01, 0.01 & {\bf82.0}   & 20.3  & 94.8   & 20.2  & 96.5   & 22.6  & 95.6   & 20.2 \\
  $n_j=50$        &       &       & 0.20, 0.05 & {\bf87.4}   & 48.1  & {\bf87.4}   & 47.0  & {92.4}   & 53.1  & {91.0}   & 51.1 \\
          &       &       & 0.20, 0.20 & {\bf87.2}   & 48.4  & {\bf87.8}   & 47.0  & 92.6   & 53.1  & {90.9}   & 51.1 \\
          &       &       & 0.60, 0.01 &{\bf89.9}   & 116.1 & {\bf87.5}  & 107.3 & {91.1}   & 121.5 & {90.9}   & 120.2 \\
          &       & .30,.10 & 0.01, 0.01 & {\bf83.2}   & 20.3  & 94.8   & 21.8  &{\it 97.2}   & 25.4  & 95.3   & 21.7 \\
          &       &       & 0.20, 0.05 & {\bf86.6}   & 47.9  & {\bf87.2}   & 46.3  & {92.1}   & 54.3  & {90.3}   & 51.2 \\
          &       &       & 0.20, 0.20 & {\bf87.7}   & 48.2  & {\bf87.2}  & 46.4  &{92.4}   & 54.1  & {90.2}   & 51.0 \\
          &       &       & 0.60, 0.01 &{\bf 89.0}   & 115.7 & {\bf85.3}   & 103.7 & {91.0}   & 121.5 &{90.7}   & 120.0 \\
          & High  & .20,.20 & 0.01, 0.01 & {\bf75.6}   & 19.4  & 94.8   & 19.8  & 96.1   & 21.9  & 94.1   & 20.0 \\
          &       &       & 0.20, 0.05 & {\bf85.4}   & 47.3  & {\bf88.9}   & 47.9  &{92.0}   & 52.7  & {91.4}   & 51.3 \\
          &       &       & 0.20, 0.20 &{\bf 85.3}   & 47.5  & {\bf88.9}   & 47.9  & {92.1}   & 52.6  & {91.4}   & 51.3 \\
          &       &       & 0.60, 0.01 &{\bf 89.7}   & 115.0 & {\bf87.9}   & 110.2 & {91.2}   & 121.0 & {91.0}  & 120.2 \\
          &       & .30,.10 & 0.01, 0.01 & {\bf74.3}   & 20.0  & 94.6   & 20.9  &{\it 97.1}   & 27.0  & 95.1   & 21.0 \\
          &       &       & 0.20, 0.05 & {\bf85.1}  & 47.6  &{\bf 87.6}   & 47.7  & 92.8   & 54.3  &{90.1}   & 51.7 \\
          &       &       & 0.20, 0.20 & {\bf84.9}   & 47.8  & {\bf87.5}   & 47.7  & {92.4}   & 53.8  & {90.2}   & 51.6 \\
          &       &       & 0.60, 0.01 & {\bf89.1}   & 115.0 & {\bf86.9}   & 108.4 & {91.0}   & 123.7 & {90.8}   & 120.9 \\ \hline
    $J=30$,   & Low   & .20,.20 & 0.01, 0.01 & {\bf81.0}   & 17.7  & 93.9   & 17.0  & {\it97.4}   & 21.0  & 93.2   & 16.9 \\
     unbalanced       &       &       & 0.20, 0.05 &{\bf 85.9}  & 32.1  &{\bf 89.9}   & 30.3  & 95.7   & 36.1  & 93.0   & 32.9 \\
          &       &       & 0.20, 0.20 &{\bf 85.9}  & 32.1  &{\bf 89.7}   & 30.3  & 96.0   & 35.9  & 93.1   & 32.7 \\
          &       &       & 0.60, 0.01 &{91.2}   & 70.3  & {\bf89.6}   & 63.7  & 94.3   & 74.0  & 93.5   & 72.4 \\
          &       & .30,.10 & 0.01, 0.01 & {\bf82.6}   & 17.8  & 93.8   & 18.4  & {\it 98.5}   & 24.6  & 93.9   & 18.3 \\
          &       &       & 0.20, 0.05 & {\bf85.7}   & 32.1  & {\bf88.0}  & 30.4  & 96.4   & 38.3  & {92.2}   & 33.4 \\
          &       &       & 0.20, 0.20 &{\bf 85.5} & 32.1  & {\bf88.2}  & 30.4  & 96.2   & 38.1  & {91.5}   & 33.1 \\
          &       &       & 0.60, 0.01 & {91.9}  & 70.5  & {\bf87.4}   & 62.1  & 94.3   & 75.1  & 94.0   & 72.9 \\
          & High  & .20,.20 & 0.01, 0.01 & {\bf74.9}   & 16.6  & 93.5   & 16.7  & {\it97.6}   & 20.4  & 93.6   & 16.8 \\
          &       &       & 0.20, 0.05 &{\bf 83.0}   & 31.1  & {90.9}   & 30.8  & 96.0   & 35.6  & 92.9   & 33.0 \\
          &       &       & 0.20, 0.20 &{\bf 83.0}   & 31.2  &{90.6}   & 30.7  & 95.9   & 35.5  & 92.8   & 32.8 \\
          &       &       & 0.60, 0.01 &{91.0}   & 69.5  &{\bf 89.7}   & 65.5  & 94.6   & 73.6  & 93.8   & 72.4 \\
          &       & .30,.10 & 0.01, 0.01 & {\bf73.4}   & 17.1  & 94.3   & 17.5  & {\it97.4}   & 22.9  & 93.8   & 17.7 \\
          &       &       & 0.20, 0.05 & {\bf82.5}   & 31.4  & {90.7}   & 30.8  & 96.0   & 37.3  & 93.5   & 33.4 \\
          &       &       & 0.20, 0.20 & {\bf82.1}   & 31.4  & {91.0}  & 30.8  & 96.2   & 36.9  & 93.0   & 33.1 \\
          &       &       & 0.60, 0.01 & {90.5}   & 69.7  & {\bf89.7}  & 64.5  & 94.0   & 74.4  & 93.8   & 72.8 \\ \hline
    \end{tabular}%

}
\end{table}%

\clearpage

\end{document}